\documentclass[aps,preprint]{revtex4}
\usepackage{amsfonts}
\usepackage{amsmath}
\usepackage{amssymb}
\usepackage{graphicx}
\usepackage{pdfpages}
\usepackage{cleveref}
\usepackage{subcaption}

\providecommand{\U}[1]{\protect\rule{.1in}{.1in}}

\begin{document}

\title{Greybody Factors of Holographic Superconductors with $z=2$ Lifshitz
Scaling}
\author{Huriye G\"{u}rsel}
\email{huriye.gursel@emu.edu.tr}
\author{\.{I}zzet Sakall{\i}}
\email{izzet.sakalli@emu.edu.tr}
\affiliation{Physics Department, Arts and Sciences Faculty, Eastern Mediterranean University, Famagusta, Northern
Cyprus, Mersin 10, Turkey}
\date{\today }
\date[Date text]{date}

\begin{abstract}
We study the quasinormal modes and thermal radiation of massless spin-0
field perturbations in the background of four-dimensional (4D) non-Abelian charged Lifshitz
black branes with $z=2$ hyperscaling violation, which correspond to systems
with superconducting fluctuations. After having an analytical solution to the
Klein-Gordon equation, we obtain exact quasinormal modes that are purely
imaginary. Therefore, there is no oscillatory behavior in the perturbations
that guarantees the mode stability of these solutions. We also study the
greybody factors, absorption cross-section, and decay rate of the
non-Abelian charged Lifshitz black branes. We derive their analytical
expressions and then investigate the correspondence in the strongly coupled dual theory. This study might shed light on the mechanism governing the
high-temperature superconductors in condensed matter physics.
\end{abstract}

\volumeyear{year}
\volumenumber{number}
\issuenumber{number}
\eid{identifier}
\date{\today }
\date[Date text]{date}
\received[Received text]{date}
\revised[Revised text]{date}
\accepted[Accepted text]{date}
\published[Published text]{date}
\maketitle
\tableofcontents

%\begin{document} 
%\maketitle
%\flushbottom

\section{Introduction}

For a wide range of physicists looking at the universe through the eyes of
experts in certain fields, anti-de Sitter/condensed matter (CM) theory (AdS/CMT)
correspondence \cite{Mald,1} appears (also known as holographic CM
physics) to be an appealing subject to dive into. One of the main reasons
for this is the fact that AdS/CMT correspondence acts as a bridge between
gravitational backgrounds, quantum field theory, and CM
physics. Throughout holographic CM physics, non-quasiparticle
transport is studied based on experimental phenomena and is compared with
black hole and black brane (BB) solutions in large field theories \cite{2}. In 1973,
Bardeen, Carter, and Hawking \cite{3} showed that black holes obey the laws of
thermodynamics; and in the sequel Hawking \cite{HR,HR2,HR3} stated that black holes
are actually not black.

\bigskip

In the years since dynamic critical exponent $z$ and
hyperscaling violation parameter $\theta $ have been proposed,
a great many research has been conducted by scientists of many professions,
both from observational and theoretical aspects \cite{Hertz,Fisher}. Throughout Ref. \cite{Hertz},
different quantum systems are studied in detail; and consequently, the
values that $z$ is allowed to possess are figured out for each case. In
addition to being regarded as the main source where the dynamic critical
exponent was suggested for the first time, Ref. \cite{Hertz} carries a vital
importance in literature, as it provides a linkage between the zero
temperature or low temperature behavior of quantum mechanical systems and
the associated $z$ values. On the other hand, Ref. \cite{Fisher} is devoted
to maintaining arguments on re-normalization and scaling, once a second
order transition controlled by a zero-temperature fixed point is achieved
for Random-Field Ising systems; and yet the significance of hyperscaling
violation parameter for such systems is pointed out. Both studies \cite{Hertz,Fisher} are supported by experimental phenomena \cite%
{Vil,Maher,Bray} and a great deal of studies are carried out emphasizing the
role of these exponents in CM physics \cite{Metlit,S1,Subir}
and in string theory \cite{string,string2}. For a detailed study on the
superconducting nature of the boundary theory and investigation of the
effect of dynamical exponent on the formation of scalar condensate, the reader is referred to \cite%
{ch}. On the observational side, mainly antiferromagnets situated in fields
are investigated in \cite{Cowley,Belanger,Birgen}; and the aforementioned
remarks are achieved. From the perspective of general relativity, however,
exploring the effect of these exponents on the wave dynamics of some
specific spacetime structures remains intact. 
\bigskip

The study of quantum fields propagating in the curved spacetime background predicts particle emission and the thermal black body spectrum is represented by the Hawking temperature \cite%
{HR,HR2,HR3,ex1,ex2,ex3,ex4,ex5,ex6,ex7,ex8,ex9,ex10,ex11,ex12}. This black body spectrum has its characteristic rate of absorption probability, absorption cross-section, and decay
rate which are all frequency-dependent quantities \cite{GF1,GF2,GF3,GF4,GF5,GF6,GF7,GF8,GEX1,GEX2,GEX3,GEX3n,GEX4,GEX5,GEX6,GEX7,GEX8,GEX9,GEX10,GEX11}. The quasinormal modes (QNMs), on the other hand, disclose how quick
a thermal state in the boundary theory will reach thermal equilibrium
according to the AdS/CFT (conformal field theory) correspondence \cite{Maldac}. This happens because the
relaxation time of a thermal state is inversely proportional to the
imaginary part of the QNMs of the dual gravity background that
was achieved by the QNMs of the bulk spacetime, which appears from the
poles of the retarded correlation function of the corresponding
perturbations of the dual CFT \cite{Birmi}. For details on QNMs and stability, one may see the following
studies \cite{esco,Becar,an1,an2,an3,an4,an5}. It is also interesting seeing the linkage between QNMs and the hydrodynamic equations of motion which are conservation laws for the only long-lived excitations expected in a strongly interacting system  \cite{Erd}. The AdS/CFT correspondence makes it possible to derive hydrodynamics from perturbations of bulk spacetime as well as transport coefficients like viscosity, conductivity and resistivity. For experimental results, see for example \cite{Kous,Yusuke,Stewart,Thomas}.
\bigskip

Throughout this work, we have
taken the opportunity to investigate the influence of $z$ and $\theta $ on
the radiation spectrum, in particular on the the greybody factor (GF) of the
$4D$ Lifshitz-like BB. In the literature, there exist
numerous studies on hyperscaling violating metrics among which some can be
viewed from \cite{Dong,Narayan,Fan,H.Lu,Deh,SourceofBecar,Becar,Pedraza}. 
Our work differs from previous studies in the manner that we
particularly focus on the analytical 
computation of GF, absorption cross-section, decay
rate, and QNMs for the charged $4D$ Lifshitz-like spacetimes having the dynamic
critical and hyperscaling violation exponents $z=2$ and $\theta =-1$,
respectively. This non-relativistic case is rather substantial, as systems of $z=2$
exhibit superconducting fluctuations \cite{Hertz}. It is worth noting that
the discovery of superconductivity dates back to 1911 and it
has managed to sustain outrageous interest since then \cite%
{Onnes,Meissner,Casimir,London1,London2,Heisenberg,Iso,Frh,Pippard,6,Ev,7,Koppe,cern}. Despite being actively used in many different fields of science,
the phase transitions of these systems still have mysteries awaiting to be
solved. In fact, many of the important properties of superconductors such as
Meissner effect and Abrikosov vortices depend on the dynamics of the field. For obtaining more
information on holographic superconductor models, one may refer to \cite{snew}
and \cite{s2} and references therein. In the study of \cite{Philip}, it is shown that Lifshitz-like
planar spacetimes could indeed be good candidates for revealing significant
information regarding the holographic superconductivity.  Remarkably, Manikandan and Jordan \cite%
{Andreev} have recently revealed a mapping between quantum physics of black holes
and thermodynamic properties of superconductors. Similarities between the
two phenomena are also presented in detail in \cite{Oxford}. As mentioned in Ref. \cite{Sh}, gravitational systems with $z=2$ can be mapped onto specific CM structures including magnetic materials, liquid crystals, and more specifically; cuprate superconductors. The AdS/CFT correspondence suggests that the boundary value of the bulk field gives a background source for the corresponding dual field theory operator $\mathcal{O}$ \cite{Hartn}. At this point, we shall remark that non-relativistic CFTs relevant to nature are strongly coupled and it is a rather challenging task to approach such systems with the usual perturbative techniques. For studies regarding non-relativistic systems, one may refer to \cite{ek1,ek2,ek3,ek4,ek5,ek6,ek7,ek8,ek9}. Furthermore, as also stated in Ref. \cite{yeni}, the experimental results on many-body properties of non-relativistic CFTs are most commonly supported via numerous numerical simulations and one can get an insight of these observational outcomes via the relevant simulations. However, it is of value noting that constructing analytical methods for those observables of the CFTs remains almost untouched in literature. With the will of finding analytical solutions for the thermal radiation parameters from semi-classical calculations of the
$4D$ Lifshitz-like BB, our pursuit is to address this literature gap via the mapping of the correspondence. 
\bigskip

The structure of the paper is as follows. Section II includes some details on the
geometrical structure of the $4D$ non-Abelian charged Lifshitz spacetimes with $z=2$
hyperscaling violation and yet provides the steps for evaluating an analytic
solution for the Lifshitz-like BBs under massless scalar perturbation in a clear
manner. Then, the massless Klein-Gordon equation (KGE) is analytically solved in
this background and the obtained solution is discussed around near
horizon and asymptotic regions. Section III includes the computations
of QNMs and stability analysis. In Sect. IV; GF, absorption
cross-section, and decay rate of the concerned BB are analytically computed from the perspective of semi-classical gravitational theory, whereas Sect. V touches upon the dual field
theory applications. The last section is devoted to summary and conclusions.

\section{Behavior of Massless Scalar Field in Non-Abelian Charged
Lifshitz Spacetime with $z=2$ Hyperscaling Violation}

\bigskip

\subsection{Geometric Structure}

For strongly-coupled systems in holographic CM physics, 't
Hooft matrix large $N$ limit needs to be taken into account. The fields, $\Phi _{k}$, of
concern are large $N\times N$ matrices and the interactions
are illustrated in \cite{1} as 
\begin{equation}
\mathcal{O}=tr(\Phi _{{k}_{1}}\Phi _{{k}_{2}}...\Phi _{{k}_{m}}),
\end{equation}%
where $k=1,2,...,N$. The Lagrangian that characterizes the dynamics of such
a system is defined as 
\begin{equation}
\mathcal{L}=\frac{N}{\lambda }tr(\partial ^{\mu }\Phi \partial _{\mu }\Phi
+...),
\end{equation}%
in which $\lambda $ denotes 't Hooft coupling and for the cases when $%
\lambda $ is large, strong interactions occur. In our scenario, there exists
a strong coupling between Einstein gravity, the cosmological constant $%
\Lambda $, and the fields of concern; namely the dilaton, Maxwell, and $N$ $%
SU(2)$ Yang - Mills fields which are denoted as $\phi $, $\mathcal{A}$ and $%
A_{k}^{a}$, respectively (note that $a$ also runs from 1 to $N$). Lifshitz
spacetime with hyperscaling violation are solutions to the Lagrangian \cite{SourceofBecar} 
\begin{equation}
\mathcal{L}=\sqrt{-g}\left[ R-V\left( \phi \right) -\frac{1}{2}\left(
\partial \phi \right) ^{2}-\overset{N}{\underset{k=1}{\sum }}\frac{1}{%
4g_{k}^{2}}e^{\lambda \phi }F_{k}^{2}-\frac{1}{4}e^{\lambda \phi }F_{\mu \nu
}^{a}F^{a\mu \nu }\right] ,  \label{LAG}
\end{equation}%
with $V(\phi )=\Lambda e^{-\lambda \phi }$ and $\Lambda =-[D(z-1)^{2}+z-1]$.
Equation (\ref{LAG}) yields the following line-element: 
\begin{equation}
ds^{2}=r^{\theta }\left( -r^{2z}f(r)dt^{2}+\frac{dr^{2}}{r^{2}f(r)}+r^{2}%
\overset{D-2}{\underset{i=1}{\sum }}dx_{i}^{2}\right) ,  \label{metric}
\end{equation}%
where 
\begin{equation}
f(r)=1-\frac{q^{2}r^{2\left( 1-z\right) }}{2\left( z-1\right) },
\label{func}
\end{equation}%
and 
\begin{equation}
\theta =\frac{2}{D-2}[z-(D-1)],  \label{th}
\end{equation}%
at which $q$ stands for the exact electric charge of concern, $g_{k}$ is
linked to coupling of the Yang Mills term, and $R$ is the Ricci scalar.
Furthermore, one shall write $F_{\mu \nu }^{a}=\partial _{\mu }A_{\upsilon
}^{a}-\partial _{\upsilon }A_{\mu }^{a}+\epsilon ^{abc}A_{\mu
}^{b}A_{\upsilon }^{c}$. \bigskip

Holographic correspondence states that the action involves fields
propagating on a higher dimensional curved spacetime \cite{SourceofBecar}. \bigskip

\subsection{Massless Scalar Wave Equation}

Since our focus in this work concerns massless scalar particles, we employ
the KGE:
\begin{equation}
\frac{1}{\sqrt{-g}}\partial _{\mu }\left( \sqrt{-g}g^{\mu \upsilon }\partial
_{\nu }\varphi \right) =0,  \label{KG}
\end{equation}%
where $\varphi $ denotes the massless scalar field. By considering the
symmetries of metric (\ref{metric}), one may set \cite{Becar} 
\begin{equation}
\varphi (t,r,\vec{x})=\Phi(r)e^{i\vec{\kappa}\cdot \vec{x}}e^{-i\omega
t},  \label{ANSATZ}
\end{equation}%
which leads to a more compact form of the KGE that will in turn be used for
evaluating the analytical radial solution. Note that $\vec{\kappa}$ and $%
\vec{x}$ represent $(D-2)$-dimensional wave and spatial vectors,
respectively, whereas $\omega $ denotes frequency of the emitted radiation.
After making straight forward computations, one can derive the generic
radial equation of Eq. (\ref{KG})\ for the metric (\ref{metric}) as 
\begin{equation}
\frac{d}{dr}\left[ f(r)r^{2+\tilde{\eta} -\tilde{\theta}}\frac{d\Phi}{dr}\right] +\frac{%
1}{r^{2+\tilde{\theta}-\tilde{\eta} }}\left( \frac{\omega ^{2}}{r^{2\left(
z-1\right) }f(r)}-\kappa ^{2}\right) \Phi(r)=0\,,  \label{grad}
\end{equation}%
in which $\tilde{\eta} =\frac{\tilde{\theta}D}{2}+z+D-3$ and $-\kappa ^{2}$ denotes
the eigenvalue of the Laplacian in the flat base submanifold \cite{Becar}.
Furthermore, setting 
\begin{equation}
\Phi(r)=\mathcal{F}(r)r^{-\xi }\,,
\end{equation}%
where $\xi =\frac{(D-2)(2+\tilde{\theta})}{4}$, and by defining the tortoise
coordinate $r^{\ast }$ \cite{CHANDRA} as 
\begin{equation}
r^{\ast }=\int r^{-\left( 1+z\right) }\frac{dr}{f(r)}\,,  \label{ast}
\end{equation}%
one can express the radial equation (\ref{grad}) as a one-dimensional Schr%
\"{o}dinger like equation (or the so-called Zerilli equation \cite{CHANDRA}) 
\begin{equation}
\frac{d^{2}\mathcal{F}(r^{\ast })}{dr^{\ast 2}}-\mathcal{V}(r)\mathcal{F}%
(r^{\ast })=-\omega ^{2}\mathcal{F}(r^{\ast }),  \label{zerilli}
\end{equation}%
where $\mathcal{V}(r)$ denotes the effective potential: 
\begin{equation}
\mathcal{V}(r)=r^{2\left( z-1\right) }f(r)\left[ \frac{q^{2}}{2}\xi
r^{3-z}+\xi (\xi +z)r^{2}f(r)+\kappa ^{2}\right] .  \label{genpot}
\end{equation}

During this study, based on our current literature knowledge, we have
seen that it is not possible to obtain the \textit{exact analytical solution} of the
generic radial equation (\ref{grad}) due to its transcendental form. As already being mentioned in the introduction, throughout this study, we consider the specific case of $z=2$ and $D=4$; henceforth $\theta =-1$ and $\xi =\frac{1}{2}$. Therefore, metric (\ref{metric}) reduces to 
\begin{equation}
ds^{2}=-H(r)dt^{2}+\frac{dr^{2}}{H(r)}+r\overset{2}{\underset{i=1}{\sum }}%
dx_{i}^{2},  \label{4dmetric}
\end{equation}%
where $H(r)=r^{3}f(r)$ and $f(r)=1-\frac{q^{2}}{2r^{2}}$. For the $4D$
non-Abelian charged Lifshitz BB (\ref{4dmetric}), the surface gravity \cite%
{wald} can be computed as follows
\begin{equation}
\kappa _{s}=\left. \frac{H^{\prime }}{2}\right\vert _{r=r_{H}}=r_{H}^{2},
\end{equation}%
noting that the outer event horizon obeys $r_{H}^{2}=q^{2}/2$. The generic
radial equation (\ref{grad}) then reduces to 
\begin{equation}
H(r)\frac{d^{2}\Phi}{dr^{2}}+(4r^{2}-2r_{H}^{2})\frac{d\Phi}{dr}+\left( \frac{%
\omega ^{2}}{H(r)}-\frac{\kappa ^{2}}{r}\right) \Phi(r)=0,  \label{radial1}
\end{equation}%
where $-\kappa ^{2}$ denotes the eigenvalue of the Laplacian in the flat base submanifold 
\cite{Becar}. Changing the variable via $\Tilde{z}=r^{-2}(r^{2}-r_{H}^{2})$ and setting 
\begin{equation}
\Phi(\Tilde{z})=\Tilde{z}^{\alpha }(1-\Tilde{z})^{\beta }G(\Tilde{z}),
\label{trans}
\end{equation}

with $\beta =3/2,$ one gets 
\begin{equation}
\Tilde{z}(1-\Tilde{z})G^{\prime \prime }(\Tilde{z})+\left( 1-\frac{7\Tilde{z}%
}{2}-\frac{i\omega (1-\Tilde{z})}{r_{H}^{2}}\right) G^{\prime }(\Tilde{z})+%
\left[ \frac{5i\omega -6r_{H}^{2}-\kappa ^{2}}{4r_{H}^{2}}\right] G(\Tilde{z}%
)=0,  \label{radial}
\end{equation}%
where $^{\prime }$ represents the derivative with respect to $%
\Tilde{z}$. Comparing Eq. (\ref{radial}) with the
hypergeometric differential equation \cite{hyper} 
\begin{equation}
\Tilde{z}(1-\Tilde{z})G^{\prime \prime }(\Tilde{z})+\left[ c-\left(
1+a+b\right) y\right] G^{\prime }(\Tilde{z})-abG(\Tilde{z})=0
\label{radial2}
\end{equation}%
results in%
\begin{equation}
G(\Tilde{z})=C_{1}\hspace{0.1cm}{}_{2}F_{1}\left( a,b;c;\Tilde{z}\right) +%
C_{2}\Tilde{z}^{1-c} \hspace{0.1cm}{}_{2}F_{1}\left(
a-c+1,b-c+1;2-c;\Tilde{z}\right) ,  \label{iz1}
\end{equation}%
with the relevant constants 
\begin{equation}
a=\alpha +\frac{5}{4}\mp \frac{\sqrt{\kappa _{s}^{2}-4\kappa _{s}\kappa
^{2}-4\omega ^{2}}}{4\kappa _{s}},  \label{gen}
\end{equation}%
\begin{equation}
b=\alpha +\frac{5}{4}\pm \frac{\sqrt{\kappa _{s}^{2}-4\kappa _{s}\kappa
^{2}-4\omega ^{2}}}{4\kappa _{s}},  \label{COEF}
\end{equation}%
\begin{equation}
c=1+2\alpha ,  \label{ccc}
\end{equation}%
where $\alpha =\pm \frac{i\omega }{2\kappa _{s}}$. Throughout this work,
without loss of generality, we choose

\begin{equation*}
\alpha =-(i\omega /2\kappa _{s}),
\end{equation*}
\begin{equation}
a=\frac{5}{4}-\frac{i}{2\kappa _{s}}\left( \omega +\widehat{\omega }\right) ,
\label{e}
\end{equation}%
\begin{equation}
b=\frac{5}{4}-\frac{i}{2\kappa _{s}}\left( \omega -\widehat{\omega }\right) ,
\label{f}
\end{equation}%
where%
\begin{equation}
\widehat{\omega }=\sqrt{\omega ^{2}+\kappa _{s}\left( \kappa ^{2}-\frac{%
\kappa _{s}}{4}\right) }.  \label{o}
\end{equation}

Thus, Eq. (\ref{ccc})\ becomes 
\begin{equation}
c=1-\frac{i\omega }{\kappa _{s}}.  \label{v}
\end{equation}

Then, the general solution for the radial function is obtained as%
\begin{equation}
\Phi(\Tilde{z})=\Tilde{z}^{\alpha }(1-\Tilde{z})^{\beta }\left[ C_{1}\hspace{%
0.1cm}{}_{2}F_{1}\left( a,b;c;\Tilde{z}\right) +C_{2}\hspace{0.1cm}\Tilde{z}%
^{1-c}\hspace{0.1cm}{}_{2}F_{1}\left( a-c+1,b-c+1;2-c;%
\Tilde{z}\right) \right] .  \label{R}
\end{equation}%
After this point, one shall split the problem into two parts and investigate
the behavior of Eq. (\ref{R}) near the event horizon and at the spatial
infinity regime separately. This will then provide the desired information
regarding the flux computation.

\subsubsection{Radial Solution Around Near Horizon Region}

To consider the near-horizon property of the solution (\ref{R}), we reconsider
the tortoise coordinate for the metric \eqref{radial1}: 
\begin{equation}
r^{\ast }=\int \frac{dr}{H(r)}.  \label{tr1}
\end{equation}

After some algebra, one can obtain the near-horizon tortoise coordinate in
terms of $\Tilde{z}$ [recall that $\Tilde{z}=r^{-2}(r^{2}-r_{H}^{2})$]\ as
follows

\begin{equation}
r_{NH}^{\ast }=\frac{\ln {\sqrt{1-\Tilde{z}}-1}}{2r_{H}^{2}}.  \label{tr2}
\end{equation}

For the case when $r\rightarrow r_{H}$, or equivalently for $\Tilde{z}%
\rightarrow 0$, the hypergeometric function becomes identity $[_{2}F_{1}\left(a,b,c;0\right)=1] $ and the general radial solution (\ref{R}) can be
expressed as 
\begin{equation}
\Phi_{NH}=C_{1}e^{\alpha \ln \Tilde{z}}+C_{2}e^{-\alpha \ln \Tilde{z}},
\label{NHs}
\end{equation}

\bigskip This enables us to rewrite Eq. (\ref{NHs}) as 
\begin{equation}
\Phi_{NH}\equiv \Phi(r\rightarrow r_{H})=C_{1}e^{-i\omega \frac{\ln \Tilde{z}}{%
2r_{H}^{2}}}+C_{2}e^{i\omega \frac{\ln \Tilde{z}}{2r_{H}^{2}}}=\widetilde{%
C_{1}}e^{-i\omega r_{NH}^{\ast }}+\widetilde{C_{2}}e^{i\omega r_{NH}^{\ast
}},  \label{NHs2}
\end{equation}

\bigskip where $\widetilde{C_{1}}=C_{1}e^{\omega \pi /2r_{H}^{2}}$ and $%
\widetilde{C_{2}}=C_{2}e^{\omega \pi /2r_{H}^{2}}.$ This implies that the
scalar field (\ref{ANSATZ}) near horizon region can explicitly be stated as%
\begin{equation}
\varphi _{NH}=C_{1}e^{-i\omega \left( t+\frac{\ln \Tilde{z}}{2r_{H}^{2}}%
\right) }=\widetilde{C_{1}}e^{-i\omega r_{NH}^{\ast }}e^{-i\omega t}.
\label{NHs4}
\end{equation}

It is clear from Eq. (\ref{NHs4}) that the first term corresponds the
ingoing wave while the second term is the outgoing wave. In order to match
the ingoing boundary condition near the horizon, the coefficient $C_{2}$
must be vanished. Then, we have the general radial solution with the ingoing
boundary condition at the horizon as 
\begin{equation}
\Phi(\widetilde{z})=C_{1}\widetilde{z}^{\alpha }(1-\widetilde{z})^{\beta
}\hspace{%
0.1cm}{}_{2}F_{1}\left( a,b;c;\Tilde{z}\right).  \label{NHs3}
\end{equation}

\subsubsection{Radial Solution Around Spatial Infinity Region}

We now turn our focus to the computation of the emitted radiation's flux at
spatial infinity. Although there exist a variety of ways for this
evaluation, we will be following the method used in \cite{Das} which
requires finding the asymptotic solution for Eq. (\ref{radial1}) followed by
performing 
\begin{equation}
\digamma _{SI}=\frac{\sqrt{-g}g^{rr}}{2i}(\Phi_{SI}^{\ast }\partial
_{r}\Phi_{SI}-\Phi_{SI}\partial _{r}\Phi_{SI}^{\ast }).  \label{myflux}
\end{equation}%
As stated in section II, our Lagrangian involves strong coupling which
implies that in order for the AdS/CMT correspondence to hold true, the low
energy GF should be of interest. Furthermore, our choice of the parameter $%
\beta $ (i.e., $\beta =3/2$) also supports this requirement, as $\beta $ being
real makes it a challenging task to distinguish between the ingoing and
outgoing fluxes \cite{Das}. Thus, for $r\rightarrow \infty $, Eq. (\ref%
{radial1}) reduces to 
\begin{equation}
\frac{d^{2}\Phi}{dr^{2}}+\frac{4}{r}\frac{d\Phi}{dr}=0,  \label{radial3}
\end{equation}%
which allows us to express the radial solution at spatial infinity as 
\begin{equation}
\Phi(r)=D_{1}+\frac{D_{2}}{r^{3}}.  \label{r33}
\end{equation}%
Having obtained the asymptotic radial solution, we will now solve Eq. (\ref%
{R}) for $r\rightarrow \infty $ and compare our solution with the one
obtained above so as to be able to find the relevant constants. Hence, it is
worthwhile to note that near the spatial infinity $\Tilde{z}\rightarrow 1$
or $r\rightarrow \infty $, the general radial solution (\ref{R}) behaves as follows%
\begin{multline}
\Phi_{SI}(\Tilde{z})=C_{1}\Tilde{z}^{\alpha }\left[A_{1} (1-\Tilde{z})^{\beta
}\hspace{%
0.1cm}{}_{2}F_{1}\left( a,b;a+b-c+1;1-\Tilde{z}\right)\right.+ \\
\left. A_{2}\left( \Tilde{z}\right)\hspace{%
0.1cm}{}_{2}F_{1}\left( c-a,c-b;c-a-b+1;1-\Tilde{z}\right)\right] .  \label{cgf}
\end{multline}

To this end, the following linear transformation relationship is employed : 
\begin{equation}
_{2}F_{1}(a,b,c;u)=A_{1}\hspace{0.1cm}{}_{2}F_{1}(a,b,a+b-c+1;1-u)+A_{2}(1-u)^{c-a-b}\hspace{0.1cm}{}_{2}F_{1}(c-a,c-b,c-a-b+1;1-u),
\end{equation}

where 
\begin{equation}
A_{1}=\frac{\Gamma (c)\Gamma (c-a-b)}{\Gamma (c-a)\Gamma (c-b)},  \label{bir}
\end{equation}
\begin{equation}
A_{2}=\frac{\Gamma (c)\Gamma (a+b-c)}{\Gamma (a)\Gamma (b)}.  \label{iki}
\end{equation}

\bigskip Thus, near the spatial infinity $\widetilde{z}\rightarrow 1$ or $%
r\rightarrow \infty $, the asymptotic behavior of the radial solution (\ref%
{cgf}) behaves as 
\begin{equation}
\Phi_{SI}=C_{1}\left[ A_{1}\left( \frac{r_{H}}{r}\right) ^{3}+A_{2}\right] .
\label{E}
\end{equation}

Matching Eq. (\ref{r33}) with Eq. (\ref{E}) results in $D_{1}=A_{2}C_{1}$ and 
$D_{2}=A_{1}C_{1}r_{H}^{3}$. Finally, the asymptotic flux (\ref{myflux})
becomes \cite{Das} 
\begin{equation}
\digamma _{SI}=3\left( \left\vert D_{out}\right\vert ^{2}-\left\vert
D_{in}\right\vert ^{2}\right) ,  \label{ddv}
\end{equation}%
in which 
\begin{equation}
D_{out}=\frac{D_{1}+iD_{2}}{2},
\end{equation}%
and 
\begin{equation}
D_{in}=\frac{D_{1}-iD_{2}}{2}.
\end{equation}

\section{QNM\MakeLowercase{s} and Stability Analysis}

In this section, we shall compute the QNMs by using the
analytical radial solutions obtained in Sec. II and analyze the stability
of the $4D$ non-Abelian charged Lifshitz BBs with $z=2$ hyperscaling
violation under the scalar field perturbation. QNMs describe
perturbations of a field that decay in time. In other words, they are the
modes of energy dissipation of a perturbed field.

For QNM analysis, it is necessary to examine the behavior of
the effective potential \cite{CHANDRA} that the wave will be subjected to.
Taking cognizance of Eq. (\ref{genpot}), one can see that the effective
potential of the $4D$ non-Abelian charged Lifshitz spacetime with $z=2$
hyperscaling violation reads 
\begin{equation}
\mathcal{V}(r)=H(r)\left( \frac{5r}{4}-\frac{r_{H}^{2}}{4r}+\frac{\kappa ^{2}%
}{r}\right) .  \label{pot}
\end{equation}%
\begin{figure}[h]
\begin{center}
\includegraphics[scale=0.4]{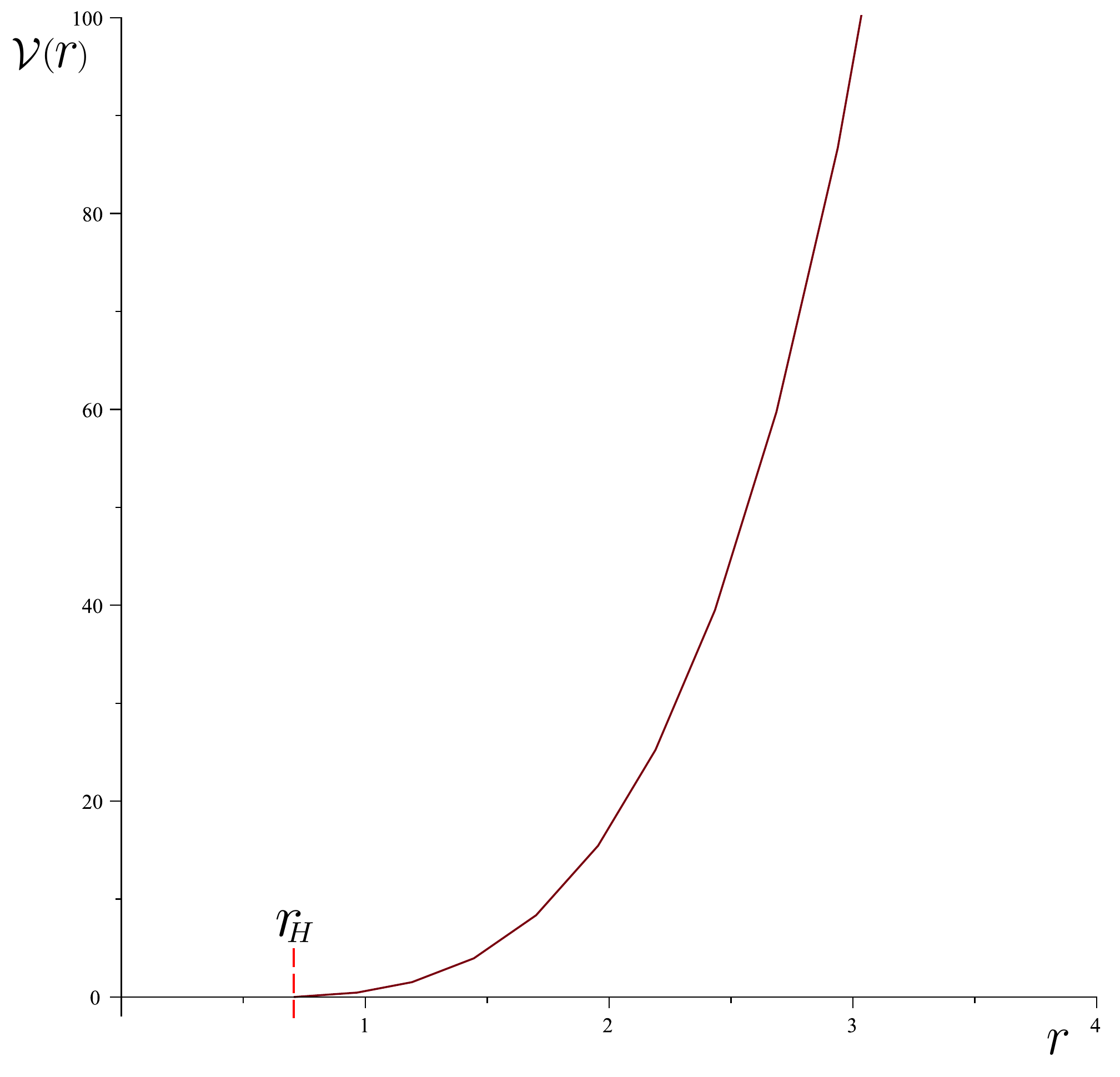}
\end{center}
\caption{The behavior of $\mathcal{V}(r)$ with $q=1$ and $\protect\kappa =0$%
. }
\label{Potential}
\end{figure}
One can check that $\underset{r\rightarrow \infty }{\lim }\mathcal{V}(r){%
\rightarrow }\infty $: this can be best seen from Fig. (1). Therefore, the QNMs possess the particular boundary
conditions such that scalar field $\varphi $ is purely ingoing at the
horizon and vanishes at spatial infinity (a similar situation was discussed
in, for example, \cite{qnmdef}). Since the asymptotic ($\Tilde{z}\rightarrow
1$) radial function~is already obtained in Eq. (\ref{cgf}), we thus have%
\begin{eqnarray}
\Phi_{SI}(\Tilde{z}) &\approx &C_{1}A_{1}(1-\Tilde{z})^{\beta }+C_{1}A_{2}, 
\notag \\
&\cong &C_{1}A_{2}=C_{1}\frac{\Gamma (c)\Gamma (a+b-c)}{\Gamma (a)\Gamma (b)}%
.  \label{sip}
\end{eqnarray}%
Therefore, the field at spatial infinity vanishes if $a=-n$ or $b=-n$ for $%
n=0,1,2,..$. The latter remarks give us the explicit expression for the
QNMs: 
\begin{equation}
\omega =-i\frac{q^{2}(n+1)(2n+3)+\kappa ^{2}}{5+4n}.  \label{w1}
\end{equation}%
Since the obtained QNMs are purely imaginary and negative, it guarantees that the system is always overdamped. Thus, one concludes that
the $4D$ non-Abelian charged Lifshitz BBs with $z=2$ hyperscaling violation
are stable under the massless scalar field perturbations.

\section{Thermal Radiation}

In this section, we will be carrying out final steps for obtaining
analytical expressions for the main focusing point of our study, the thermal
radiation parameters $\gamma$, $\sigma_{abs}$, and $\Gamma$, which stand for
the GF, absorption cross-section, and decay rate, respectively. To obtain
these parameters, we need to start from the GF evaluation, which is
described in \cite{Das} as follows
\begin{equation}
\gamma=1 - \Re= \frac{2i(\mathfrak{D}-\mathfrak{D}^{\ast})}{\mathfrak{D}%
\mathfrak{D}^{\ast}+i(\mathfrak{D}-\mathfrak{D}^{\ast})+1},  \label{GF}
\end{equation}
where $\Re=\left\vert D_{out}\right\vert^{2}/\left\vert
D_{in}\right\vert^{2} $ and $\mathfrak{D}=D_{1}/D_{2}$. More precisely, we have
\begin{equation}
\mathfrak{D}=\frac{3}{8}\frac{\Gamma\left(-\frac{1}{4} - i
X\right)\Gamma\left(-\frac{1}{4}- i Y \right)}{\Gamma\left(\frac{5}{4}- i
Y\right)\Gamma\left(\frac{5}{4} - i X\right)r_{H}^{3}},
\end{equation}
and hence
\begin{equation}
\mathfrak{D} \mathfrak{D}^{\ast}=\frac{2304}{\pi^{4}r_{H}^{6}}\hspace{0.1cm}[\Gamma(3/4)]^{8} \hspace{0.1cm} {\displaystyle%
\prod\limits_{n=0}^{\infty}} \frac{\varepsilon_{n}}{\left[ 1+ \left(\frac{Y}{%
n-1/4}\right)^{2}\right] \left[ 1+ \left(\frac{X}{n-1/4}\right)^{2}\right]},
\end{equation}
\begin{equation}
\mathfrak{D}-\mathfrak{D}^{\ast}=\frac{24 }{\pi^{2}r_{H}^{3}} \hspace{0.1cm}\Xi 
\hspace{0.1cm} [\Gamma(3/4)]^{4}{\displaystyle\prod\limits_{n=0}^{\infty}} 
\hspace{0.1cm} \varepsilon_{n}.
\end{equation}
Note that the simplifications above are achieved with the aid of the
following relations: 
\begin{equation}
X=\frac{\omega - \hat{\omega}}{2r_{H}^{2}},
\end{equation}
\begin{equation}
Y=\frac{\omega + \hat{\omega}}{2r_{H}^{2}},
\end{equation}
\begin{equation}
\varepsilon_{n}= \left[ 1+ \left(\frac{Y}{n+5/4}\right)^{2}\right] \left[ 1+
\left(\frac{X}{n+5/4}\right)^{2}\right],
\end{equation}
and 
\begin{equation}
\Xi=\frac{(\sin{\theta_{1}}\sin{\theta_{2}} - \sin{\theta_{3}} \sin{%
\theta_{4}})}{\sin{\theta_{1}}\sin{\theta_{2}}\sin{\theta_{3}}\sin{\theta_{4}%
}}.
\end{equation}
Furthermore, the associated angles can be defined as $\theta_{1}=\pi\left(%
\frac{5}{4}-iX\right)$, $\theta_{2}=\pi\left(\frac{5}{4}-iY\right)$, $%
\theta_{3}=\pi\left(\frac{5}{4}+iY\right)$, and $\theta_{4}=\pi\left(\frac{5}{%
4}+iX\right)$. At this point, one shall record that we have taken the
advantage of the following properties of the gamma functions throughout our
calculations: 
\begin{equation}
\frac{\Gamma(x+iy) \Gamma(x-iy)}{[\Gamma(x)]^{2}}= {\displaystyle%
\prod\limits_{n=0}^{\infty}} \left[1+ \left(\frac{y}{x+n}\right)^{2}\right]%
^{-1},
\end{equation}
and the reflection formula 
\begin{equation}
\frac{1}{\Gamma(Z)}\frac{1}{\Gamma(1-Z)}= \frac{\sin{\pi Z}}{\pi},
\end{equation}
where $Z \in \mathbb{C}$ \cite{Magnus}. The aforementioned simplifications allow us
to express the GF in anew form as follows: 
\begin{equation}
\gamma= \frac{2i \hspace{0.1cm}\Xi \hspace{0.1cm}{\displaystyle%
\prod\limits_{n=0}^{\infty}} \varepsilon_{n}}{\frac{96 \hspace{0.1cm}
[\Gamma(3/4)]^{4}}{\pi^{2}r_{H}^{3}}{\displaystyle\prod\limits_{n=0}^{\infty}} 
\hspace{0.1cm} \frac{%
\varepsilon_{n}}{\left[1+ \left(\frac{Y}{n-1/4}\right)^{2}\right] \left[ 1+
\left(\frac{X}{n-1/4}\right)^{2}\right]} + i \hspace{0.1cm} \Xi {%
\displaystyle\prod\limits_{n=0}^{\infty}} \hspace{0.1cm} \varepsilon_{n} + \frac{\pi^{2}r_{H}^{3}}{%
24[\Gamma(3/4)]^{4}}}.  \label{GF3}
\end{equation}
Having obtained the GF, one may now have the virtue of evaluating some other
thermodynamic quantities. Let us start by computing the $4D$ absorption
cross-section which reads \cite{GEX8,GEX9,GEX10}%
\begin{equation}
\sigma_{abs}=\sum_{l=0}^{\infty}\frac{i\pi}{\omega^{2}}\frac{2(2l+1) \hspace{
0.1cm}\Xi  \hspace{
0.1cm} {\displaystyle\prod\limits_{n=0}^{\infty}} \varepsilon_{n}}{\frac{96 
\hspace{0.1cm} [\Gamma(3/4)]^{4}}{\pi^{2}r_{H}^{3}} \hspace{0.1cm} {\displaystyle\prod\limits_{n=0}^{\infty}} \frac{
\varepsilon_{n}}{\left[1+ \left(\frac{Y}{n-1/4}\right)^{2}\right] \left[ 1+
\left(\frac{X}{n-1/4}\right)^{2}\right]} + i \hspace{0.1cm} \Xi \hspace{0.1cm} {\displaystyle%
\prod\limits_{n=0}^{\infty}} \varepsilon_{n} + \frac{\pi^{2}r_{H}^{3}}{%
24[\Gamma(3/4)]^{4}}}.   \label{acs}
\end{equation}
Finally, the decay rate of our concerned BB (\ref{metric}) is represented as%
\begin{equation}
\Gamma=\frac{i \hspace{0.1cm}\Xi \hspace{0.1cm}{\displaystyle%
\prod\limits_{n=0}^{\infty}} \varepsilon_{n}\hspace{0.1cm}d^{3}k}{4
\pi^{3}(e^{\omega/T_{H}}-1)\left[\frac{96 \hspace{0.1cm} [\Gamma(3/4)]^{4}}{%
\pi^{2}r_{H}^{3}} \hspace{0.1cm} {%
\displaystyle\prod\limits_{n=0}^{\infty}} \frac{\varepsilon_{n}}{\left[1+
\left(\frac{Y}{n-1/4}\right)^{2}\right] \left[ 1+ \left(\frac{X}{n-1/4}%
\right)^{2}\right]} + i \hspace{0.1cm} \Xi  {\displaystyle\prod\limits_{n=0}^{%
\infty}} \varepsilon_{n} + \frac{\pi^{2}r_{H}^{3}}{%
24[\Gamma(3/4)]^{4}}\right]}. 
\end{equation}
\bigskip

\section{Duality Between Analytical Results and Strongly Coupled CFT Systems}

In the previous sections, we have used the tools of semi-classical methods to compute QNMs, GF, absorption cross section, and decay rate of charged Lifshitz-like background with hyperscaling violation of $z=2$ and $\theta=-1$, under scalar perturbations. In the holographic scenario, the dual theory is constructed on the boundary of the bulk spacetime which is located at infinite radial distance away \cite{1105.6335}. Thus, the bulk fields in $4D$ gravitational model are directly linked  to dual operators in the dual field theory of two-spatial dimensions on the boundary. It is worth recalling that in our work, the field propagating in the curved spacetime of our concern was chosen to be massless scalar field as $\varphi(t,r,\vec{x})=\Phi(r) \hspace{0.05cm} e^{i \vec{\kappa}.\vec{x}} \hspace{0.05cm}e^{-i \omega t}$. Hence, in the boundary field theory, $\Phi$ will correspond to $\mathcal{O}_{\Phi}$, namely a scalar operator.

We will now be touching upon the relevance of our analytical gravitational results to the strongly coupled systems exhibiting quantum behavior. Let us start from the linkage between QNMs obtained in Eq. (\ref{w1}) and thermalization in the dual strongly coupled CFT. Prior to doing so, we shall take advantage of the membrane paradigm which states that small fluctuations of a stretched horizon have properties corresponding to diffusion of a conserved charge in simple fluids \cite{0309213,Thorne,Nar2}. In other words, a dispersion relation of the form $\omega=-i \mathcal{D} q^{2}$ suggests the existence of diffusion of a conserved charge \cite{0309213}. Comparing the dispersion relation with the obtained QNM (\ref{w1}), one can see that
\begin{equation}
\mathcal{D} =\frac{(n+1)(2n+3)}{5+4n}\label{Q},   \end{equation}
where $\mathcal{D}$ stands for the shear mode diffusion constant \cite{0309213,Diff}. This constant plays a significant role in AdS/CFT correspondence, as its consistency can be investigated via experimental realizations \cite{0309213,0205051}. For the fundamental QNM, i.e. for $n=0$, our diffusion constant reduces to $\mathcal{D}=3/5$. The diffusion constant has a direct relation with the inverse relaxation time. For further details on the numerical analysis of inverse relaxation times for different systems, one may refer to the study of Horowitz et. al. \cite{9909056}.
\bigskip

Having calculated the diffusion constant, let us now inspect its relation with the ratio $\eta/s$. Recall metric (\ref{metric}) for  $D=4$:
\begin{equation}
ds^{2}=r^{\theta }\left( -r^{2z}f(r)dt^{2}+\frac{dr^{2}}{r^{2}f(r)}+r^{2}%
\overset{2}{\underset{i=1}{\sum }}dx_{i}^{2}\right).
\end{equation}

Letting $r\rightarrow1/\tilde{r}$ and $\theta\rightarrow-\tilde{\theta}$ leads to

\begin{equation}
 d\tilde{s}^{2}=\tilde{r}^{\tilde{\theta}}\left( -\frac{f(\tilde{r})}{\tilde{r}^{2z}}dt^{2}+\frac{d\tilde{r}^{2}}{\tilde{r}^{2}f(\tilde{r})}+
\overset{2}{\underset{i=1}{\sum }}\frac{dx_{i}^{2}}{\tilde{r}^{2}}\right),    
\end{equation}
which coincides with the non-relativistic holographic backgrounds considered in \cite{Nar2}. For consistency, we have set the boundary spatial dimension to $d_{i}=2$. \textcolor{blue}{In Ref. \cite{Nar2}; Kolekar, Mukherjee and Narayan derived a universal relation for the shear viscosity to entropy density ratio. We will now be checking how our QNM analysis can be related to this universal relation.}
\bigskip

Firstly, we shall record that our choice of exponents ($d=D-1=3$, $z=2$ and $\tilde{\theta}=1$) satisfy the null energy conditions \cite{Nar2}
\begin{equation}
(d-1-\tilde{\theta})((d-1)(z-1)-\tilde{\theta})\geqslant0, \hspace{1cm} (z-1)(d-1+z-\tilde{\theta})\geqslant0.
\end{equation}

According to Kolekar, Mukherjee, and Narayan \cite{Nar2}, the shear viscosity to entropy density ratio obeys a \textit{universal} relation derived from the membrane paradigm 
\begin{equation}
\frac{\eta}{s}= \frac{d-z+1}{4\pi}  \hspace{0.1cm}\mathcal{D}  \hspace{0.1cm}\tilde{r}_{H}^{2-z}\label{Denk}.
\end{equation}
\textcolor{blue}{On the other hand, in another study \cite{ek8} of the same authors, it was reported that the result obtained above applies to uncharged hyperscaling violation theories and may differ for the charged backgrounds. In fact, $\eta/s$ ratio for the charged BBs is still an active debate topic \cite{ek8n}.}

\textcolor{blue}{For $d=3$, $z=2$, and $\mathcal{D}=3/5$ (i.e., for the fundamental QNMs), Eq. (\ref{Denk}) becomes}
\begin{equation}
\frac{\eta}{s}= \frac{3}{10 \pi}\label{Denkl};
\end{equation}
and yet, satisfies the so-called universal Kovtun-Son-Starinets  bound: $\eta/s\geqslant\frac{1}{4\pi}$ \cite{0405231}. \textcolor{blue}{This result carries importance both in the bulk theory and the dual CFT, as the possible experimental verification of this number from either theory would imply the following: The first implication could be that the system under experimental investigation would highly probably be exhibiting properties of bulk spacetime considered in this work. Furthermore, the experimental verification would suggest that the validity of Eq. (\ref{Denk}) proposed by Kolekar, Mukherjee and Narayan can be extended to charged hyperscaling violating Lifshitz-like backgrounds as well.} It is also worthwhile mentioning that the experimental constancy of lower bound for the $\eta/s$ ratio in CM systems is suspected to be an inherent property of semi-classical gravitational theory and should be valid for any theory with a gravitational dual description \cite{0205051}. 

 Now, let us further investigate the mapping between CM systems and their corresponding bulk spacetime models. In dual strongly coupled CFT, the two-point correlation function which corresponds to the retarded Green's function, plays a vital role, since its computation allows one to obtain exact or numerical values for physical observables like conductivity, resistivity, flux factor, cross section, shear viscosity, and so on. In order to achieve this, a well-defined boundary value for $\Phi$ should initially be computed. Then, one can follow the method prescribed by Gubser-Klebanov-Polyakov-Witten (GKPW) \cite{kitap}. According to the GKPW, we consider an infinitesimal distance $\epsilon$ away from the boundary of the bulk spacetime. Although this will lead to modifications in the relevant action, the equations of motion remain invariant. Consequently, the UV divergence is avoided and taking $\epsilon\rightarrow 0 $  results in a well-defined boundary value for $\Phi$. The flux factor then can be evaluated via \cite{Kous}
\begin{equation}
\digamma(\vec{\kappa},\omega)=\underset{r\rightarrow \epsilon}{\lim}\hspace{0.1cm} \sqrt{g}\hspace{0.1cm} g^{rr} \Phi(r)\hspace{0.1cm} \partial_{r}\Phi(r),  
\end{equation}
which corresponds to the momentum-space two-point correlation function, i.e,
\begin{equation}
\digamma(\vec{\kappa},\omega)=\left<\mathcal{O}_{\Phi}(\vec{\kappa},\omega) \hspace{0.1cm} \mathcal{O}_{\Phi}(-\vec{\kappa},-\omega)\right>.   
\end{equation}
For further details, the reader is referred to \cite{H.Lu,Sh,Kous,TEO} and references therein. This factor carries a major significance in real-world experiments. For instance, for particle physics experiments involving scattering processes, the transitions between states constitute the observables of the system and the model benefits from non-relativistic perturbation theory \cite{Halzen}. Finally, the differential cross-section can be obtained via \cite{Halzen}
\begin{equation}
d\sigma=\frac{1}{\digamma}\left\vert \mathcal{M}\right\vert^{2} d\tilde{\Phi},
\end{equation}
in which $\mathcal{M}$ and $d\tilde{\Phi}$ represent matrix element and phase factor, respectively. Now, let us inspect our case. Recalling Eq. (\ref{r33}):
\begin{equation}
\Phi(r)=D_{1}+\frac{D_{2}}{r^{3}},
\end{equation}
which represents the asymptotic behavior of our radial solution $\Phi$, one can get the retarded Green's function as $G_{O\:+}=D^{-1}$ where $D^{-1}=D_{2}/D_{1}$ \cite{H.Lu}. Then, one can express
\begin{equation}
G_{O\:+}=\frac{8}{3}\frac{\Gamma\left(\frac{5}{4}- i
Y\right)\Gamma\left(\frac{5}{4} - i X\right)r_{H}^{3}}{\Gamma\left(-\frac{1}{4} - i
X\right)\Gamma\left(-\frac{1}{4}- i Y \right)}.
\end{equation}
As one may notice, the Green's function obtained above is actually the key expression that one needs for being able to evaluate the physical observables in the theory of our concern. For instance, to be able to obtain an analytical solution for
the universal $\eta$ in hydrodynamics, one can use  \cite{1810.09242}
\begin{equation}
\eta=-\underset{\omega\rightarrow 0}{\lim}\hspace{0.1cm} \frac{1}{\omega}\hspace{0.1cm} 
Im\hspace{0.1cm} G_{O\:+}(\omega,\kappa=0),
\end{equation}
which would indeed be useful for mapping gravitational results into the dual field theory.
\bigskip
\section{Conclusion}

The main motivation behind our work was the idea of using the tools of semi-classical gravitational theory to perceive quantum behaviour of strongly-coupled systems of the physical world. Equipped with this motivation, we have evaluated the thermal radiation parameters of hyperscaling violating Lifshitz BB solutions with $z=2$ to gravity-dilaton-Maxwell-Yang-Mills theories in 4D in the bulk spacetime. Although the parameters we have evaluated do carry significance in gravitational theory, one shall note that they also have intriguing implications in non-relativistic CFTs. 
\bigskip

In this study, we have first focused on the scalar perturbations of the
non-Abelian Lifshitz spacetime with $z=2$ hyperscaling violation that have provided
us with the analytical expressions for QNMs, GF, absorption
cross-section, and decay rate of this Lifshitz BB. We have seen that the obtained exact
QNMs are purely imaginary; and one shall note that in such perturbations, an
exponential decay behavior is observed. Namely, the system is
always over-damped, which results in the mode stability of the non-Abelian
Lifshitz spacetime with $z=2$ hyperscaling violation. Having obtained the gravitational observables, we have then touched upon the linkage between these analytical results and CM systems possessing strong coupling. For the fundamental QNM, the shear viscosity to entropy density ratio in our non-relativistic model is found to be $\frac{\eta}{s}= \frac{3}{10 \pi}$,
which satisfies the universal Kovtun-Son-Starinets bound. \textcolor{blue}{Although the ratio of $\eta/s$ is an ongoing research topic for the charged hyperscaling violating theories \cite{ek8,ek8n}, we believe that the present study will provide contribution to the relevant discussions.} Finally, we have evaluated the thermal Green's function in the dual theory in terms of the exact expressions we had obtained in the semi-classical gravitational theory.
\bigskip

Our future plans include detailed evaluation of transport coefficients under AdS/CFT correspondence and seeking for relevant experimental evidence for the model of our concern. Furthermore, one can also construct an effective string configuration for verification of the calculations carried out in this work. We are planning to dive into this in the near future with the aid of Ref. \cite{Alwis}.  We will then search for comparison of our results with the numerical and experimental studies in literature. We also hope that the exact solutions obtained in this work do get experimentally verified in superconducting systems which can in turn be
used throughout gathering more information on strongly coupled fluids. The
desire of attaining the exact CM analogue of our analytic results
acts as a motivation for further research and discussion, as there exists a
broad range of applications of AdS/CMT correspondence in many different
areas of physics. Finally, it is worthwhile to re-investigate the outcomes
of this study in the presence of back-reaction. We hope to be able to report
on this case in the near future.

\section*{Acknowledgements}

The authors are grateful to the Editor and anonymous Referee for their
valuable comments and suggestions to improve the paper.

\end{document}